\documentclass[prb,aps,twocolumn,draft,tightenlines, %
showpacs,floatfix]{revtex4}
\usepackage{psfig}
\usepackage{amssymb}
\usepackage{amsmath}
\usepackage{colordvi}
\begin{document}
\title{Schottky-barrier induced spin relaxation in spin injection}
\author{Y. Y. Wang}
\author{M. W. Wu}
\thanks{Author to whom correspondence should be addressed}%
\email{mwwu@ustc.edu.cn.}
\affiliation{Hefei National Laboratory for Physical Sciences at
  Microscale,University of Science and Technology of China,
Hefei, Anhui, 230026, China}
\affiliation{Department of Physics,University of Science and Technology
of China,  Hefei, Anhui, 230026, China}
\altaffiliation{Mailing Address.}

\date{\today}
\begin{abstract}

An ensemble Monte Carlo method is used to study the spin injection through a
ferromagnet-semiconductor junction where a Schottky barrier is formed.
It is shown that the  Schottky-barrier-induced  electric field
which is confined in the depletion region and is
parallel to the injection direction,
is very large. This electric
field can induce an effective magnetic field due to the Rashba effect and
cause  strong spin relaxation.
\end{abstract}

\maketitle
Spin injection from a ferromagnetic metal contact into a
non-magnetic low-dimensional  semiconductor  structure
is one of the prerequisites for the
realization of the next generation  high-speed low-power
devices based on spin degree of freedom.\cite{wolf,spintronics,das}
Notwithstanding the fact that many efforts have been
devoted to this problem experimentally,\cite{han,ham}
an efficient room-temperature spin injection is still far away
from the horizon.\cite{ham} In the meantime, there are
many theoretical investigations \cite{theo,brat,cheng}
on the spin injection through the ferromagnet-semiconductor
junction where a high potential Schottky barrier
is formed.\cite{sze} In these studies the interface is
treated through various boundary conditions.\cite{theo,brat}
Large (up to 100\ \%) spin injections  are reported in these theories.
Very recently Shen {\em et al.} reported a first
ensemble Monte Carlo (MC) simulation of the spin
injection through a Schottky barrier into a semiconductor quantum well
(QW).\cite{cheng} In this study, the Schottky barrier is
treated carefully through the simulation.
Still they reported a substantial spin polarization
after the injection  to a length scale
in the order of 1\ $\mu$m at room temperature without external
magnetic field. Therefore there must be something missing in the theories in
dealing with the ferromagnet-semiconductor junctions.

It is noted that the  Schottky barrier induces a very large
electric field parallel to the QW.
Such an electric field can induce an effective magnetic field
due to the Rashba effect\cite{yu} and can therefore
cause a strong reduction of the spin polarization after the injection.
This effect has long been neglected
in the literature. A quantitative estimation of this new relaxation mechanism
requires an accurate computation of the electric field induced by the Schottky
barrier which varies
strongly with the position and is confined in the depletion region.
For this purpose we apply an ensemble MC
simulation to simulate the Schottky barrier and  examine the spin relaxation
induced by this additional relaxation mechanism under various
conditions.

We study a ferromagnet-semiconductor diode which is one of the
elements for many spintronic devices.\cite{datta} The
spin-polarized particles are injected from a bulk ferromagnetic
metal into a GaAs QW through a Schottky barrier by both thermionic
emission and tunnelling injection, excluding the recombination in
the space-charge region and the hole injection from the metal to
the semiconductor.\cite{sze} The direction of injection is
parallel to the QW plane. The electron transport in the QW is
based on the semiclassical approximation, simply including a
``drift'' and a ``scattering'' process: During the ``drift''
process, the spin is influenced by both the Rashba\cite{yu} and
the Dresselhaus\cite{dresselhaus} spin-orbit interactions. The method of the MC
simulation has been laid out in detail in Ref. \onlinecite{sun}
for the Schottky barrier simulation, in Refs.\
\onlinecite{bournel,shen} for the spin transport simulation and in
Ref.\ \onlinecite{cheng} for the spin injection simulation. For
the inhomogeneous electron distribution in the depletion region,
the compression/expansion technique  presented by Martin {\em et
al.}\cite{martin} has to be applied. In this report we do not repeat these details
except the differences which are addressed in the following.

At finite temperature $T$, the total current
injected from a ferromagnetic metal to a semiconductor through the
Schottky barrier is written as \cite{sze}
\begin{equation}
j_{ms}(E_{x}) = \frac{A^{\ast} T }{k_{B}} \int^{\infty}_{0} T_{ms}(E_{x})
f_{m}(E)
[1-f_{sc}(E)] dE\ ,
\label{current}
\end{equation}
where $k_{B}$ denotes the Boltzmann constant and $A^{\ast}$ stands
for the Richardson constant.  $T_{ms}(E_{x})$ is the tunnelling
probability through the barrier at the energy $E_{x}$ which
represents the kinetic energy along the $x$-direction (the
injection direction). It is $1$ for $E_{x} \geqslant\Phi_{B}$ and
$T_{ms}(E_{x})=\exp\{-\frac{2}{\hbar} \int^{x_{tp}}_{0} \sqrt{2
m^{\ast}
  [E_{c}(x)-E_{x}]}\}$
for $0< E_{x} < \Phi_{B}$ following the WKB approximation.
$\Phi_{B}$ represents the Schottky barrier height at the metal-semiconductor
 interface, $x_{tp}$ is the
electron position after tunnelling and $E_{c}(x)$ stands for the
bottom of the conduction band in the semiconductor. $f_{m}(E)$ and
$f_{sc}(E)$ are the electron distribution functions in the
ferromagnetic metal and semiconductor separately with $E$ standing
for the total energy. It is emphasized here that unlike the
previous works,\cite{cheng,sun}
 the current and the tunnelling probability are
only functions of $E_{x}$, instead of $E$. After injection,
electrons start travelling in the QW subject to the spin-orbit
interactions, the electric field and the electron-phonon and
possible electron-impurity scattering. The spin-orbit interaction
is described by $H_{so}(x) = H_{R}(x) + H_{D}$, with the spacial
variable-dependent Rashba term
\begin{equation}
H_{R}(x) = \gamma [(\sigma_{x} k_{y} -\sigma_{y} k_{x}){\cal E}_{z}-
  k_{y}\sigma_{z} {\cal E}_{x}(x)]\ ,
\label{rashba}
\end{equation}
the linear Dresselhaus term
$H_{D}^{(1)} = \beta \langle k_{z}^{2}\rangle
 (\sigma_{y}k_{y}- \sigma_{x} k_{x})$ and the cubic
Dresselhaus term $H_{D}^{(3)}=\beta (\sigma_{x} k_{x} k_{y}^{2}
 - \sigma_{y} k_{y} k_{x}^{2})$.
${\cal E}_x(x)$ in Eq.\ (\ref{rashba}) is the
Schottky-barrier-induced electric field (SBIEF). The effective magnetic
field induced by it in Eq.\ (\ref{rashba}) has a
significant influence to the spin relaxation and has long been
overlooked in the literature.

We apply the MC method to study the spin injection from
the magnetic Fe to (001) GaAs QW through a Schottky contact. The
well width is $8$\ nm. The Schottky barrier height in the
simulation is fixed to be $\Phi_{B}=0.72$\ eV.\cite{Waldrop} We
use the following spin-orbit coupling constants: $\beta = 28$\
eV$\cdot$\AA$^{3}$ for the Dresselhaus effect\cite{Cardona} and
$\gamma =740$ eV$\cdot$\AA$^{2}$ for the Rashba one.\cite{Nitta}
The channel length along the spin transport is $L_x=2.5$\ $\mu$m.
The injection takes place at the Fe/GaAs interface at $x=0$. As
we investigate the spin injection from the source,  the drain is
assumed to be in Ohmic contact with the QW. In the figures of this
paper, we only show the results for the initial $1$\ $\mu$m. In the MC
simulation, we divide $L_x$ into 500 cells and choose the time
step to be $\Delta t =1$\ fs. To achieve the steady transport
region, we run the simulation program for $10000$ time steps and
get the results by averaging over the last $3000$ steps. The
initial spin polarization is always assumed to  be along the $x$-axis
throughout this paper.

\begin{figure}[htb]
  \centerline {\psfig{figure=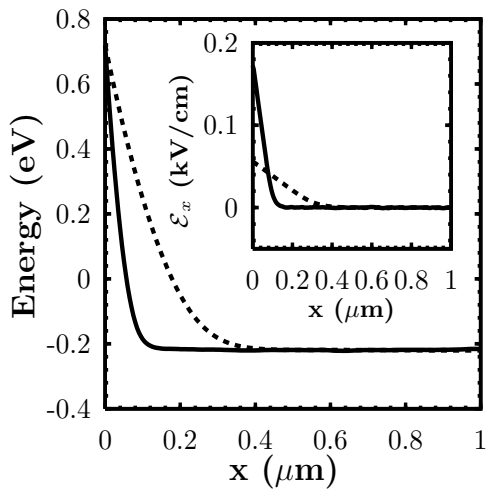,width=6.5cm,height=6.5cm,angle=0}}
\caption{Energy of the simulated Schottky barriers for two electron densities $n$
at bias $V=0.2$\ V and
  $T=300$\ K. Solid curve: $n=10^{11}$ cm$^{-2}$; Dotted  curve: $n=10^{10}$
  cm$^{-2}$. The corresponding
  electric fields are shown in the inset.}
\label{fig1}
\end{figure}

The simulated Schottky barrier shape, which is determined by the
solution of the Poisson equation and the MC simulation of the
electron distributions self-consistently, is shown in Fig.\ 1
for different electron densities $n$ in semiconductor QW. An
inverse bias voltage $V=0.2$\ V is applied in the simulation, which is in
favor of the electron injection from the ferromagnetic metal into
the semiconductors.\cite{sze} The large bending near the contact
indicates the existence of a depletion
layer where the electron concentration is negligible. It also
gives an electric field ${\cal E}_x(x)$ which is shown in the inset of
the figure.
It can be seen that the Schottky barrier becomes thicker when the
electron density in the QW is decreased. This is in consistent
with the approximation relation that the Schottky
barrier width is proportional to $n^{-1/2}$.\cite{sze} Due to
the change of the  shape of the barriers, the SBIEF changes also
at different electron densities as shown
in the insect of Fig.\ 1.

Because of the large population of the spin-unpolarized electrons in
the device, especially beyond the depletion region, the total spin
polarization averaged over all the particles at a given position
reduces to nearly zero at about $x=20$ nm.\cite{albrecht,cheng} We
want to get the spin evolution of the injected electrons, so our
simulation only get the spin polarization at each grid averaged
over the injected spin-polarized electrons. In fact, the spin
polarization of electrons in the interface of the
ferromagnetic metal is determined by the spin-dependent density of
states of electrons in the ferromagnetic contact. Nevertheless, in
order to investigate spin polarization clearly, we assume the
injected carrier is $S_{x}=100$\ \% spin polarized first. We use
$|S|=\sqrt{S_{x}^{2}+S_{y}^{2}+S_{z}^{2}}$ to denote the spin
polarization of the injected electrons. Moreover, differing from
the previous works\cite{cheng,shen} where the electron density is
as high as 10$^{12}$\ cm$^{-2}$, in the present work we only
concentrate on the case with density being smaller than 10$^{11}$\ cm$^{-2}$.
This is because that when the electron density is high, the
chemical potential is large compared to $k_{B}T$. Therefore one
should not use the Boltzmann distribution. Nevertheless, the MC
method treats the scattering semiclassically and does not contain
any distribution function. Consequently, it can only be applied to
the problems with low electron density.

\begin{figure}[htb]
  \centerline
  {\psfig{figure=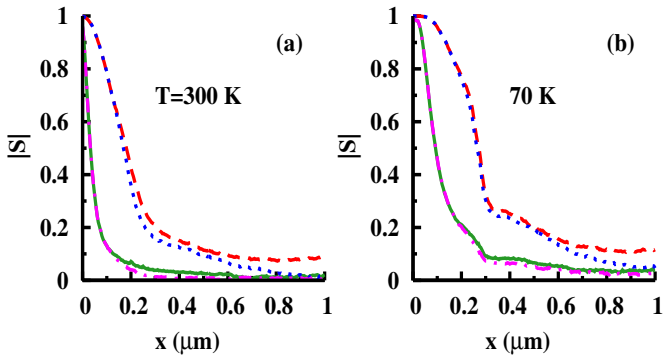,width=9.cm,height=5cm,angle=0}}
\caption{(Color online) Spin polarization evolution at different
  temperatures without (the red and blue  or the dashed and
dotted curves) and with (the green and pink or the solid and
chain curves)
SBIEF ${\cal E}_{x}(x)$ in the Rashba term. The
effect of the cubic Dresselhaus term is shown by including (the blue and
pink or the dotted and chain curves) and
excluding (the red and green or the dashed and solid curves) this term
in the simulation. $n=10^{10}$\ cm$^{-2}$.
}
\label{fig2}
\end{figure}

 In Fig.\ 2 the spin polarization $|S|$ is plotted as function of
 the position $x$ at temperatures $T=300$ and 70\ K
 without (the red and blue  or the dashed and dotted curves)
 and with (the green and pink or the solid and
chain curves) the
SBIEF ${\cal E}_x(x)$ in the Rashba term [Eq.\ (\ref{rashba})].
 The electron density is
 10$^{10}$\ cm$^{-2}$ in the simulation.
 We also show the effects of the cubic Dresselhaus term
  to the spin relaxation by performing the simulation
 with (the blue and
pink or the dotted and chain curves) and without (the red and green or
the dashed and solid curves) $H_D^{(3)}$. It is seen from the figure that
the SBIEF ${\cal E}_x(x)$ in Eq.\ (\ref{rashba}) leads to a pronounced spin
relaxation in the depletion region. The
spin polarization is almost zero after the depletion layer, in
contracts to the previous report of a substantial amount at the
length scale of 1\ $\mu$m.\cite{cheng} The spin relaxation in all
the cases is due to the D'yakonov and Perel' mechanism.\cite{yakonov}
It is further noted from the figure that after the
fast initial drop of the spin polarization in the depletion region,
the spin polarization also slowly  decreases with the position.
This is because of the spin relaxation induced by the Rasbhba terms
from the electric field perpendicular to the QW, {\em i.e.}, ${\cal E}_z$, and
the Dresselhaus terms.
This is in contrary  to the results reported by Shen {\em et
al.},\cite{cheng} where they show that the spin polarization keeps almost
constant beyond the depletion region.  It is  further seen from the
figure that the third-order Dresselhaus term has marginal effect on the
spin relaxation, especially at the depletion region where
the Rashba term is dominant.
This is because that the energy along the
$y$ direction ($k_BT$) is small, so that $k_{y}^{2}$ is small compared to
$k_{z}^{2}$ in the Dresselhaus term.

\begin{figure}[htb]
  \centerline
  {\psfig{figure=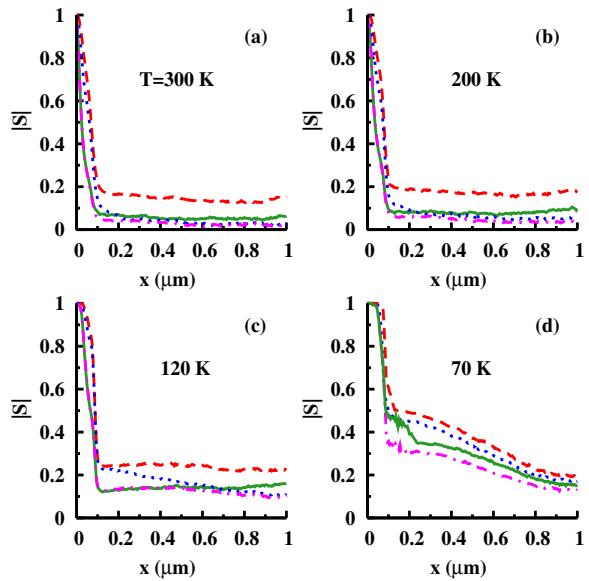,width=8cm,height=8cm,angle=0}}
\caption{(Color online) Same as Fig.\ 2 but with the electron
 density $n=10^{11}$ cm$^{-2}$.}
\label{fig3}
\end{figure}

We further investigate the effect of the SBIEF to the spin injection
at higher electron density (but still barely in the
non-degenerate regime). Curves in Fig.\ 3 are exactly
corresponding to the cases in Fig.\ 2 except the electron
density being $10^{11}$ cm$^{-2}$, an order of magnitude larger.
It is seen from the figure that after the depletion region
of Fig.\ 3, the injected spin polarizations
all become smaller compared to the corresponding cases in
Fig.\ 2. This is due to the enhanced Rashba
and Dresselhaus terms at high electron densities.
It is seen from the figure that  the effective magnetic field induced by the
SBIEF ${\cal
E}_{x}(x)$ in Eq.\ (\ref{rashba}) again markedly reduces the spin injections.
 It is further noted from the figure that the cubic Dresselhaus term
$H_D^{(3)}$ shows larger influence than the low electron density
case. This is because that the SBIEF is much lager at the
high electron density case (see Fig.\  1). This field drives electrons
 to a much larger
$|k_x|$ and gives a larger cubic
Dresselhaus term. It is also seen from Fig.\ 3 that the
effect of cubic Dresselhaus term gradually reduces with the decrease
of temperature. This is because  $k_{y}^{2}$  becomes smaller
for lower temperature.

\begin{figure}[htb]
  \centerline
  {\psfig{figure=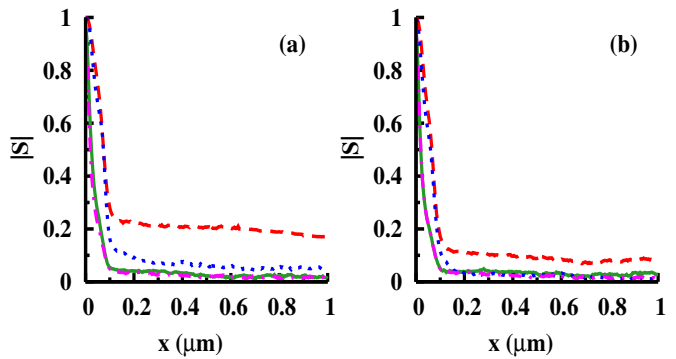,width=9cm,height=5cm,angle=0}}
\caption{(Color online) Comparison of the MC simulation with the Blotzmann sampling (a)
 and the Fermi sampling (b) for electrons at relatively high density $n=10^{11}$\ cm$^{-2}$.
$T=300$\ K. The meanings of the curves are all the same as those in Figs.\ 2 and 3.}
\label{fig4}
\end{figure}

It is noted that the MC method cannot be applied to the strong
degenerate (high density) case as
reported by Shen {\em et al.} where the electron density is taken as
high as 10$^{12}$\ cm$^{-2}$.\cite{cheng,shen} In the strong degenerate case,
the electron distribution in the scattering cannot be overlooked any more and
the MC method fails. Moreover, the
Boltzmann sampling which is independent of the density, should be changed into
Fermi sampling. In fact, even for the density at 10$^{11}$\ cm$^{-2}$,
the non-degenerate approximation is already barely valid.
In Fig.\ 4 we show the spin injections by using different samplings
(Fig.\ 4(a) for Boltzmann sampling and Fig.\ 4(b) for Fermi sampling) at
$n=10^{11}$\ cm$^{-2}$, with the scattering
still kept to be semiclassical.
One can clearly find the marked difference.
This is because when the
Boltzmann sampling is used, the energy along the $y$-axis is fixed in the
range of $k_BT$, regardless of the density. However, it is much larger
by the Fermi sampling at high density  case.
 Therefore, at high density case, the Rashba and the Dresselhaus terms
are both weaker from the Boltzmann sampling. This leads to a
larger spin polarizations. For density at 10$^{10}$\ cm$^{-2}$
reported in Fig.\ 2, both
samplings give the same results.

\begin{figure}[htb]
  \centerline
  {\psfig{figure=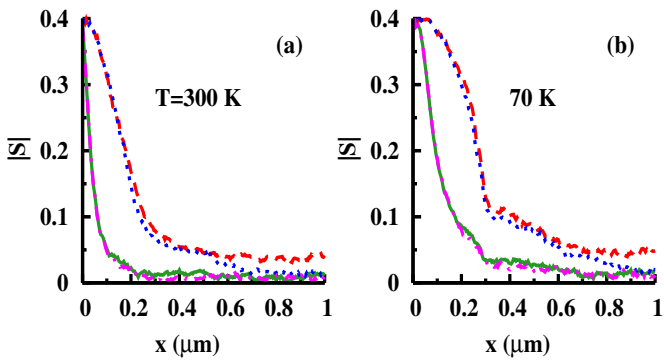,width=9cm,height=5cm,angle=0}}
\caption{(Color online) Same as Fig.\ 2 but with the initial spin
polarization being 40\ \%.
  }
\label{fig5}
\end{figure}

In Figs.\ 2-4, we assume the injected
electron spin polarization is 100\ \%. In fact, the initial spin
injection, which is determined by the spin-state probability of electrons in
the ferromagnetic contact and can be obtained from the
microscopic models of ferromagnetic
metal,\cite{butler} is about 40\ \%. Figure\ 5 shows
the results of the spin polarization inside the
semiconductor QW with electron density $n=10^{10}$\ cm$^{-2}$,
 when the initial injected electron spin polarization is 40\ \%.
The main results are all the same as those shown in Fig.\ 2.
As the initial spin polarization
is low, the spin polarization beyond the depletion layer is lower
than the corresponding case in Fig.\ 2 and reduces to zero
more quickly.

In conclusion, an ensemble MC method is used to simulate the
spin-polarized electron
injection through a Schottky barrier and transport in 2D
semiconductor QW with both Rashba and Dresselhaus spin-orbit coupling.
We show that the SBIEF not only drives electron to a higher momentum states
during the injection, which influences the spin
relaxation via the Dresselhaus
and the Rashba term, but also provides an effective magnetic field
due to the Rashba effect. We show that this SBIEF-induced effective magnetic field
is very strong and gives a pronounced effect to the spin dephasing at the
Schottky barrier area.
Consequently the spin injection becomes almost negligible after the Schottky barrier
region. This effect has long been overlooked in the literature.
Moreover, this effective magnetic field also provides additional
relaxation due to the
many-body effect\cite{wu} which is beyond the scope of the MC
simulation and will be reported elsewhere.

This work was supported by the Natural Science Foundation of China
under Grant Nos. 90303012 and 10247002, the Natural Science Foundation
of Anhui Province under Grant No. 050460203 and SRFDP.
We would like to thank S. T. Chui, M.-C. Cheng, L. Sun, M. Shen and
S. Saikin for
valuable discussions. One of the authors (MWW) would like to thank S. Zhang at
University of Missouri Columbia for hospitality where this paper is finalized.


\begin{thebibliography}{10}
\bibitem{wolf} S. A. Wolf, J. Supercond. {\bf 13}, 195 (2000).
\bibitem{spintronics} {\it Semiconductor spintronics and quantum
    computation}, ed. by D. D. Awschalom, D. Loss, and N. Samarth
    (Springer-Verlag, Berlin, 2002).
\bibitem{das} I. \v Zuti\'c, J. Fabian, and S. Das Sarma,
Rev. Mod. Phys. {\bf 76}, 323 (2004).
\bibitem{han}A. T. Hanbicki, B. T. Jonker, G. Itskos, G. Kioseoglou, and
A. Petrou, Appl Phys. Lett. {\bf 80}, 1240 (2002);
V. F. Motsnyi, J. De Boeck, J. Das, W. Van Roy, G. Borghs, E. Goovaerts, V. I.
Safarov, {\em ibid.} {\bf 81}, 265 (2002);
 A. T. Hanbicki, O. M.
J. van't Erve, R. Magno, G. Kioseoglou, C. H. Li, abd B. T. Jonker,
{\it ibid}. {\bf 82}, 4092 (2003).
\bibitem{ham}P. R. Hammer, B. R. Bennett, M. J. Yang, and M. Johnson,
Phys. Rev. Lett. {\bf 83}, 203 (1999); H. J. Zhu, M.
 Ramsteiner, H. Kostial, M. Wassermeier, H. P. Sch\"onherr, and K. H.
Ploog, {\it ibid.} {\bf 87},
016601 (2001); T. Manago and H. Akinaga, Appl. Phys. Lett. {\bf 81}, 694
(2002); H. Ohno, K. Yoh, K. Sueoka, K. Mukasa, A. Kawaharazuka, and
M. E. Ramesteiner, Jpn. J. Appl. Phys., Part 1 {\bf 42}, L1 (2003).
\bibitem{sze}S. M. Sze, {\it Physics of Semiconductor Devices} (Wiley, New
York, 1981).
\bibitem{theo}M. Johnson and J. Byers, Phys. Rev. B {\bf 67}, 125112 (2003);
G. Schmidt, D. Ferrand, L. W. Molenkamp, A. T. Filip, and B. J. van
Wees, Phys. Rev. B {\bf 62}, R4790 (2000);
E. I. Rashba, {\it ibid.} {\bf 62}, R16267 (2000); A. Fert and H. Jaffr\`es,
{\em ibid.} {\bf 64}, 184420 (2001);
 J. D. Albrecht and D. L. Smith,
{\em ibid}. {\bf 66}, 113303 (2002)
\bibitem{brat} A. M. Bratkovsky and V. V. Osipov,
J. Appl. Phys. {\bf 96}, 4525 (2004); V. V. Osipov and A. M. Bratkovsky,
Phys. Rev. B {\bf 70}, 205312 (2004).
\bibitem{cheng}M. Shen, S. Saikin, and M.-C. Cheng, J. Appl. Phys. {\bf 96},
4319 (2004).
\bibitem{yu}Yu. Bychkov and E. I. Rashba, J. Phy. C. {\bf 17}, 6039
  (1984).
\bibitem{datta}S. Datta and B. Das, Appl. Phys. Lett. {\bf 56}, 665
  (1990).
\bibitem{dresselhaus}G. Dresselhaus, Phys. Rev. {\bf 100}, 580 (1955).
\bibitem{sun}L. Sun, X. Y. Liu, M. Liu, G. Du, and R. Q. Han,
  Semicond. Sci. Technol. {\bf 18}, 576 (2003).
\bibitem{bournel}A. Bournel, V. Delmouly, P. Dollfus, G. Tremblay, and
  P. Hesto, Physica E {\bf 10}, 86 (2001).
\bibitem{shen}S. Saikin, M. Shen, M.-C. Cheng, and V. Privman,
  J. Appl. Phys. {\bf 94}, 1769 (2003);
M. Shen, S. Saikin, M.-C. Cheng, and V. Privman, Math. Comput. Simul.
{\bf 65}, 351 (2004).
\bibitem{martin}M. J. Mart\'{i}n, T. Gonzalez, D. Pardo, and
  J. E. Velazquez, Semicond. Sci. Technol. {\bf 11}, 380 (1996).
\bibitem{Waldrop}J. R. Waldrop, Appl. Phys. Lett. {\bf 44}, 1002
  (1984).
\bibitem{Cardona}M. Cardona, N. E. Christensen, and G. Fasol,
  Phys. Rev. B {\bf 38}, 1806 (1988).
\bibitem{Nitta}J. Nitta, T. Akazaki, and H. Takayanagi, Phys. Rev. B
  {\bf 38} 1806 (1988).
\bibitem{albrecht}J. D. Albrecht and D. L. Smith, Phys. Rev. B {\bf
    68}, 035340 (2003).
\bibitem{yakonov}M. I. D'yakonov and V. I. Perel', Sov, Phys. JETP {\bf
  33},1053 (1971).
\bibitem{butler}W. H. Butler, X. G. Zhang, Xindong Wang, Jan Van EK,
  and J. M. MacLaren, J. Appl. Phys. {\bf 81}, 5518 (1997).
\bibitem{wu} M. W. Wu and C. Z. Ning, Eur. Phys. J. B {\bf 18}, 373 (2000);
M. Q. Weng and M. W. Wu, J. Appl. Phys. {\bf 93}, 410 (2003); Phys. Rev. B
{\bf 68}, 075312 (2003);  M.Q. Weng, M.W. Wu, and L.
Jiang, {\em ibid.} {\bf 69}, 245320 (2004).
\end{thebibliography}
\end{document}